\documentclass[aps,prb,reprint,groupedaddress]{revtex4-1}

\usepackage{textcomp,bm}
\usepackage[]{graphicx}
\usepackage{dcolumn}
\usepackage{amssymb}   
\usepackage{amsmath}
\usepackage{ulem}

\begin{document}

\preprint{}

\title{Coupling between heavy fermion superconductor CeCoIn$_5$ and antiferromagnetic metal CeIn$_3$ through the atomic interface}


\author{M. Naritsuka}
\email[]{naritsuka@scphys.kyoto-u.ac.jp}
\author{S. Nakamura}
\author{Y. Kasahara}
\author{T. Terashima}
\author{R. Peters}
\author{Y. Matsuda}
\email[]{matsuda@scphys.kyoto-u.ac.jp}
\affiliation{Department of Physics, Kyoto University, Kyoto 606-8502, Japan}

\date{\today}

\begin{abstract}

 To study the mutual interaction between unconventional superconductivity and magnetic order through an interface,  we fabricate hybrid Kondo superlattices consisting of alternating layers of  the heavy-fermion superconductor  CeCoIn$_5$ and the antiferromagnetic (AFM) heavy-fermion metal CeIn$_3$.    
The strength of the AFM fluctuations is tuned by applying hydrostatic pressure to the CeCoIn$_5$($m$)/CeIn$_3(n)$ superlattices with $m$ and $n$ unit-cell-thick layers of CeCoIn$_5$ and CeIn$_3$, respectively.    The superconductivity in CeCoIn$_5$ and the AFM order in CeIn$_3$ coexist in spatially separated layers in the whole thickness and  pressure ranges.   
At ambient pressure, the N\'{e}el   temperature $T_N$ of the CeIn$_3$ block layers (BLs) of CeCoIn$_5$(7)/CeIn$_3(n)$  shows little dependence on the thickness $n$, in sharp contrast to CeIn$_3(n)$/LaIn$_3(4)$ superlattices where $T_N$ is strongly suppressed with decreasing $n$.   This suggests that each  CeIn$_3$ BL  is magnetically coupled by the RKKY interaction through the adjacent  CeCoIn$_5$ BL and a three-dimensional magnetic state is formed.
With applying pressure to CeCoIn$_5$(7)/CeIn$_3(13)$, $T_N$ of the CeIn$_3$ BLs is suppressed up to 2.4\,GPa, showing a similar pressure dependence as bulk CeIn$_3$ single crystals.   An  analysis of the upper critical field reveals that the  superconductivity in the CeCoIn$_5$ BLs is barely influenced by the AFM fluctuations in the CeIn$_3$ BLs, even when the CeIn$_3$ BLs are in the vicinity of the AFM quantum critical point.  This is in stark contrast to CeCoIn$_5$/CeRhIn$_5$ superlattices, in which the superconductivity in the CeCoIn$_5$ BLs is profoundly affected by AFM fluctuations in the CeRhIn$_5$ BLs.  The present results show that although  AFM fluctuations are injected  into the CeCoIn$_5$ BLs from the CeIn$_3$ BLs through the interface, they barely affect the force which binds superconducting electron pairs.  These results demonstrate  that two-dimensional AFM fluctuations are essentially important for the pairing interactions in CeCoIn$_5$.  

\end{abstract}

\maketitle

\section{Introduction}

 It is well established that in several compound families,  such as high-$T_c$ cuprates, iron pnictides/chalcogenides, and heavy-fermion compounds, Cooper pairs are not bound together through phonon exchange but instead through exchange of some other kind, such as spin fluctuations\cite{Sigrist,Bennemann,Thalmeier,Scalapino,Stewart2017,Keimer2015,Hirschfeld,Kontani}. Despite tremendous efforts, however, the interplay between unconventional superconductivity and magnetism still remains largely unexplored  in these systems.  This includes fascinating electronic phases, where superconductivity and antiferromagnetic (AFM)  order,  involving the same charge carriers, coexists, and the important question why superconductivity is often strongest near a quantum critical point  (QCP) where the AFM order vanishes in the zero temperature limit and spin fluctuations become singular \cite{Mathur1998,Shibauchi2014,Hashimoto2012,Knebel2008,Park2006}.


 By using a recent state-of-the-art molecular beam epitaxy (MBE) technique,  we grow artificial Kondo superlattices with alternating layers of heavy-fermion superconductors and conventional metals or heavy-fermion AFM compounds \cite{Shishido2010, Shimozawa2016}.    These Kondo superlattices provide unique opportunity to study  the mutual interactions between the unconventional superconducting state and magnetically ordered-  or conventional metallic-states through the atomic interface and thereby seek answers to the above-mentioned questions.   Until now,  several types of Kondo superlattices containing the heavy-fermion superconductor CeCoIn$_5$ \cite{Petrovic2001} with a layered structure have been fabricated \cite{Mizukami2011, Goh2012,Shimozawa2014, Ishii2016,Naritsuka2017, Naritsuka2018}.  CeCoIn$_5$ has a quasi-two dimensional (2D) Fermi surface\cite{Settai2007} and the presence of quasi-2D AFM fluctuations has been reported in the normal state \cite{Kawasaki2003, Raymond2015}.   
 Furthermore, a superconducting gap with $d_{x^2-y^2}$-wave symmetry has been observed by a variety of experiments \cite{Izawa2001, An2010,Matsuda2006,Stock2008,Zhou2013,Zhou2013, Allan2013}.  The superconducting state is strongly Pauli limited, as demonstrated by a first-order phase transition at upper critical fields for directions parallel and perpendicular to the $ab$ plane \cite{Izawa2001, Tayama2002,Bianchi2002, Shimahara}.  It is a prototypical system, in which  non-Fermi liquid behaviors in the normal state and unconventional superconductivity are thought to arise from the proximity to a AFM QCP \cite{Sidorov2002,Nakajima2007,Sarro2007}.   
  Under pressure, CeCoIn$_5$ moves away from the QCP and Fermi liquid behavior is recovered. 
  
  \begin{figure}[t]
	\begin{center}
		\includegraphics[width=1\linewidth]{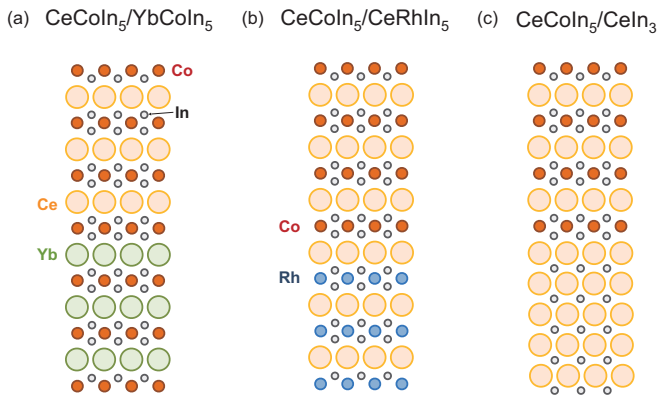}
		\caption{Schematic representations of three types of Kondo superlattices, (a) CeCoIn$_5$/YbCoIn$_5$ (b) CeCoIn$_5$/CeRhIn$_5$, and (c) CeCoIn$_5$/CeIn$_3$, where CeCoIn$_5$ is a heavy-fermion $d$-wave superconductor, YbCoIn$_5$ is a conventional metal, and CeRhIn$_5$ and CeIn$_3$ are heavy-fermion AFM metals. The atomic views of the [100] plane are shown. }
	\end{center}
\end{figure}
  
 It has been shown that in superlattices consisting of alternating layers of CeCoIn$_5$ and the  conventional metal YbCoIn$_5$ with atomic layer thicknesses (Fig.\,1a), the Pauli pair-breaking effect is strongly suppressed from that in the bulk of CeCoIn$_5$ single crystals\cite{Goh2012,Shimozawa2014}.  
  Site-selective nuclear magnetic resonance (NMR) measurements on CeCoIn$_5$/YbCoIn$_5$ superlattices have reported that  AFM fluctuations in the CeCoIn$_5$ block layers (BLs), particularly in the vicinity of the interface, are weakened\cite{Yamanaka2015}.  These results have been attributed to the local inversion symmetry breaking at the interface, which results in spin-split Fermi surfaces and thus effectively suppresses the Zeeman effect\cite{Goh2012,Maruyama2012,Shimozawa2014}.  
  
  In superlattices consisting of alternating layers of CeCoIn$_5$ and the heavy-fermion AFM metal CeRhIn$_5$ (Fig.\,1b), the superconducting- and AFM-states coexist in spatially separated layers. 
     In these superlattices, the influence of the  local inversion symmetry breaking at the interface has been shown to be less important compared to CeCoIn$_5$/YbCoIn$_5$.   In sharp contrast to  CeCoIn$_5$/YbCoIn$_5$, NMR measurements have revealed that  magnetic fluctuations in CeCoIn$_5$ BLs of  CeCoIn$_5/$CeRhIn$_5$ superlattices are enhanced compared to bulk CeCoIn$_5$ single crystal, highlighting the importance of the magnetic proximity effect \cite{Nakamine2019}.  
    In particular, it has been pointed out that in the vicinity of the QCP of CeRhIn$_5$ BLs, AFM  fluctuations are enhanced and  the force binding superconducting electron-pairs acquires an extremely strong-coupling nature.  This indicates that superconducting pairing can be manipulated by magnetic fluctuations injected through the interface \cite{Naritsuka2018}.   
  
To obtain further insight into the mutual interactions between unconventional superconductivity and  magnetic order,  we here fabricate superlattices consisting of alternating layers of CeCoIn$_5$ and the AFM metal CeIn$_3$ (Fig.\,1c). CeIn$_3$ is an isotropic Kondo lattice material with cubic crystal structure.  
  In bulk CeIn$_3$ single crystals, AFM  order with ordered magnetic moment of 0.48\,$\mu_B$ occurs  
at $T_N$=10\,K, where $\mu_B$ is the Bohr magneton \cite{Benoit1980}.   With applying pressure, $T_N$ decreases and vanishes at $\sim$2.6\,GPa, indicating an AFM QCP.  Superconductivity with a maximum $T_c\approx$200\,mK is induced in a very narrow pressure range around the QCP \cite{Mathur1998,Knebel2001}.   

Our results reveal that, similar to CeCoIn$_5$/CeRhIn$_5$ but in contrast to CeCoIn$_5$/YbCoIn$_5$ superlattices,  the local inversion symmetry breaking at the interface  has only little effect on the superconductivity in CeCoIn$_5$/CeIn$_3$ superlattices.  However, we find that the magnetic and the superconducting properties in CeCoIn$_5$/CeIn$_3$ are in marked contrast to those in CeCoIn$_5$/CeRhIn$_5$ superlattices\cite{Naritsuka2018}.   Although the AFM fluctuations are injected to the  CeCoIn$_5$ BLs from the CeIn$_3$ BLs through the interfaces, they barely affect the electron pairing interactions in the CeCoIn$_5$ BLs.  These results provide compelling evidence that  2D AFM fluctuations are essentially important for the superconductivity in CeCoIn$_5$.

\section{Experimental Details}

\begin{figure}[t]
	\begin{center}
		\includegraphics[width=0.85\linewidth]{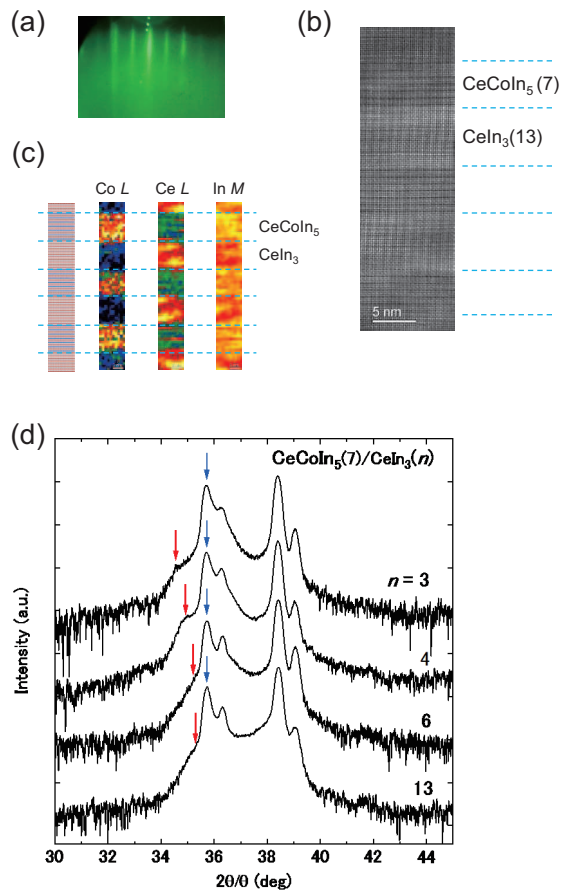}
		\caption{(a) Typical RHEED streak patterns for CeCoIn$_5$(7)/CoIn$_3$(13) superlattice taken during the crystal growth. (b), (c) High-resolution cross-sectional (b) TEM image and (c) EELS images for the CeCoIn$_5$(7)/CoIn$_3$(13) superlattice with the electron beam alined along the (100) direction. The EELS images were taken for Co $L$, Ce $L$, and In $M$ edges. (d) Cu $K\alpha_1$ x-ray diffraction patterns for CeCoIn$_5$(7)/CoIn$_3$($n$) superlattices ($n=3$, 4, 6, and 13). The blue and red arrows indicate the [003] peaks of CeCoIn$_5$ and satellite peaks due to the superlattice structure, respectively. }
	\end{center}
\end{figure}

The hybrid superlattices CeCoIn$_5$(7)/CeIn$_3$($n$) ($n$=3, 4, 6 and 13) with $c$ axis oriented structure are grown on a MgF$_2$ substrate by MBE technique \cite{Shishido2010,Shimozawa2016}.    We first grow $\sim$20\,unit-cell-thick (UCT) CeIn$_3$ ($\sim$10 nm) as a buffer layer on MgF$_2$. Then 7-UCT CeCoIn$_5$ and $n$-UCT CeIn$_3$ ($n$=3, 4, 6 and 13) are grown alternatively with total  thicknesses of approximately 200\,nm.   As the epitaxial growth temperature of CeCoIn$_5$ and CeIn$_3$ layers are different,  CeCoIn$_5$ and CeIn$_3$ BLs were grown at 570 and 420\,\textdegree{}C, respectively.  The superlattice is capped with $\sim$5\,nm Co to prevent oxidation. Streak pattern of the reflection high-energy electron diffraction (RHEED) image shown in Fig.\,2(a) have been observed during the whole growth of the superlattices, indicating good epitaxy.  The atomic force microscope measurements reveal that the surface roughness is within $\pm$1\,nm, which is comparable to 1-2 UCT along the $c$ axis of the constituents. 
Because atomically flat regions extend over distances of $\sim$0.1\,$\mu$m,  it can be expected that  transport properties are not seriously influenced by the roughness.  Figure\,2(b) displays a high-resolution cross-sectional transmission electron microscope (TEM) image along the (100) direction for the CeCoIn$_5$(7)/CeIn$_3$(13) superlattice.  A clear interface between the CeCoIn$_5$ and the CeIn$_3$ layers is observed.  Figure\,2(c) displays an electron energy loss spectroscopy (EELS) image of the same superlattice. The EELS  images clearly resolve the 7-UCT CeCoIn$_5$ and the 13-UCT CeIn$_3$ BLs, demonstrating sharp interfaces with no atomic interdiffusion between the neighboring CeCoIn$_5$ and CeIn$_3$ BLs.  Figure\,2(d) shows the X-ray diffraction patterns for CeCoIn$_5$/CeIn$_3$($n$) superlattices. The  shoulder structure shown by the red arrows near the [003] peak of CeCoIn$_5$ (blue arrows)  is consistent with the superlattice structure. 
These results demonstrate the successful fabrication of epitaxial superlattices with sharp interfaces. High-pressure resistivity measurements have been performed under  pressure up to 2.4\,GPa using a piston cylinder cell with Daphne oil 7373 as the pressure transmitting medium. The pressure has been measured by the $T_c$ of Pb. 

\begin{figure}[t]
	\begin{center}
		\includegraphics[width=1\linewidth]{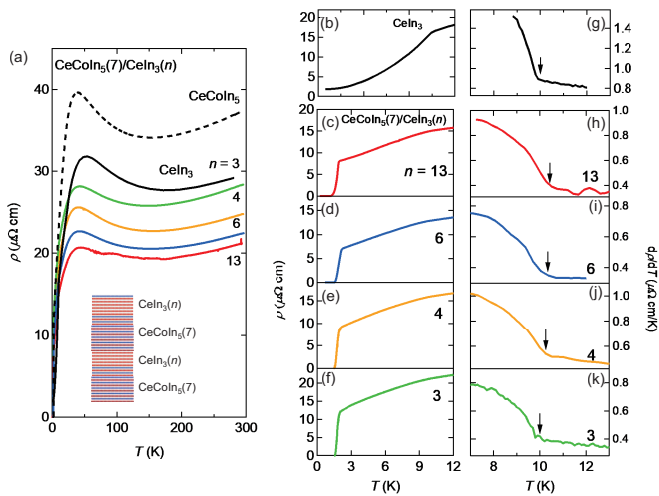}
		\caption{(a) Temperature dependence of the resistivity $\rho(T)$ in CeCoIn$_5$(7)/CeIn$_3$($n$) superlattices for $n=$3, 4, 6, and 13, along with $\rho(T)$ for CeIn$_3$ and CeCoIn$_5$ thin films. Inset illustrates the schematics of CeCoIn$_5$(7)/CeIn$_3$($n$) superlattice. (b)-(f) $\rho(T)$ at low temperatures.   (g)-(f) Temperature derivative of the resistivity, $d\rho(T)/dT$, as a function of temperature. The arrows indicate the N\'{e}el temperature $T_N$. }
	\end{center}
\end{figure}

\section{Results}

Figure\,3(a) depicts the temperature dependence of the resistivity $\rho$ of CeCoIn$_5$(7)/CeIn($n$) superlattices with $n$=3, 4, 6 and 13.  We also show $\rho$ of CeCoIn$_5$ and CeIn$_3$ thin films grown by MBE. The resistivity of  CeCoIn$_5$(7)/CeIn($n$) superlattices  follows the typical heavy-fermion behavior.  With decreasing temperature, $\rho(T)$ increases below $\sim$150\,K due to the Kondo scattering but then begins to decrease due strong $c$-$f$ hybridization between  $f$-electrons and conduction ($c$) band electrons, leading to the narrow $f$-electron band at the Fermi level.  Figures \,3(b)-3(f)  depict $\rho(T)$ at low temperatures.  All superlattices show the superconducting transition at $\approx$1.5\,K.  For the $n$=3- and 4-superlattices, $\rho(T)$ exhibit a slight downward curvature. Figures\,3(g)-3(k) display the temperature derivative of the resistivity $d\rho(T)/dT$.  As shown by the arrows in Fig.\,3(g), $d\rho(T)/dT$ of CeIn$_3$ thin film exhibits a distinct kink at $T_N$=10\,K \cite{Benoit1980}. Similar kink structures are observed in all superlattices at the temperatures indicated by arrows, showing the  AFM transition.  

\begin{figure}[t]
	\begin{center}
		\includegraphics[width=0.7\linewidth]{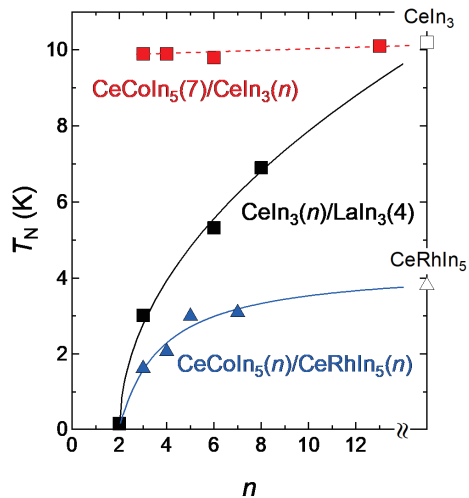}
		\caption{The N\'{e}el temperature $T_N$ for CeCoIn$_5$(7)/CeIn$_3$($n$) as a function of $n$.  For comparison,  $T_N$ for CeIn$_3$(n)/LaIn$_3$(4) and CeCoIn$_5$($n$)/CeRhIn$_5$($n$) are shown.   Open square and triangle are $T_N$ of bulk CeIn$_3$ and CeRhIn$_5$ single crystals, respectively. }
	\end{center}
\end{figure}

Figure\,4 shows the thickness dependence of $T_N$ of the CeCoIn$_5$(7)/CeIn$_3$($n$) superlattices.  For comparison, the data sets of CeIn$_3(4)$/LaIn$_3$($n$),  where LaIn$_3$ is a nonmagnetic conventional metal with no $f$-electrons \cite{Shishido2010},  and  CeCoIn$_5$($n$)/CeRhIn$_5$($n$) are also included in the figure.
Remarkably, the observed thickness dependence of $T_N$ in CeCoIn$_5$/CeIn$_3$ is in striking contrast to that in CeIn$_3$/LaIn$_3$; While $T_N$  is strongly suppressed with decreasing $n$ and vanishes at $n$=2 in CeIn$_3$/LaIn$_3$,  $T_N$ is nearly independent of $n$ in CeCoIn$_5$(7)/CeIn$_3$($n$).  
       This suggests  that  CeIn$_3$ BLs are coupled weakly by the Ruderman-Kittel-Kasuya-Yosida (RKKY) interactions through the adjacent LaIn$_3$ BL, but they can strongly couple through the adjacent CeCoIn$_5$ BL.  This is even more surprising, as the distance between different CeIn$_3$ BLs is larger in the CeCoIn$_5$(7)/CeIn$_3$($n$) superlattices than in the CeIn$_3$($n$)/LaIn$_3$(4) superlattices.
             We thus conclude that small but finite magnetic moments are induced in CeCoIn$_5$ BLs in CeCoIn$_5$/CeIn$_3$, which mediate the RKKY-interaction. On the other hand, because of the absence of strongly interacting $f$-electrons in  LaIn$_3$, which can form magnetic moments, the RKKY interaction in CeIn$_3$/LaIn$_3$ can be expected to be much weaker.
     To clarify this, a microscopic probe of magnetism, such as NMR measurements, is required.   
          We note that as shown in Fig.\,4,  the reduction of $T_N$ is also observed in CeCoIn$_5$($n$)/CeRhIn$_5$($n$) superlattices \cite{Naritsuka2018}, suggesting that the RKKY interaction between CeRhIn$_5$ BLs through adjacent CeCoIn$_5$ BL is negligibly small.   This is supported by the recent site-selective NMR measurements which report no discernible magnetic moments induced in the CeCoIn$_5$ BLs in CeCoIn$_5$/CeRhIn$_5$ \cite{Nakamine2019}.


\begin{figure}[t]
	\begin{center}
		\includegraphics[width=1\linewidth]{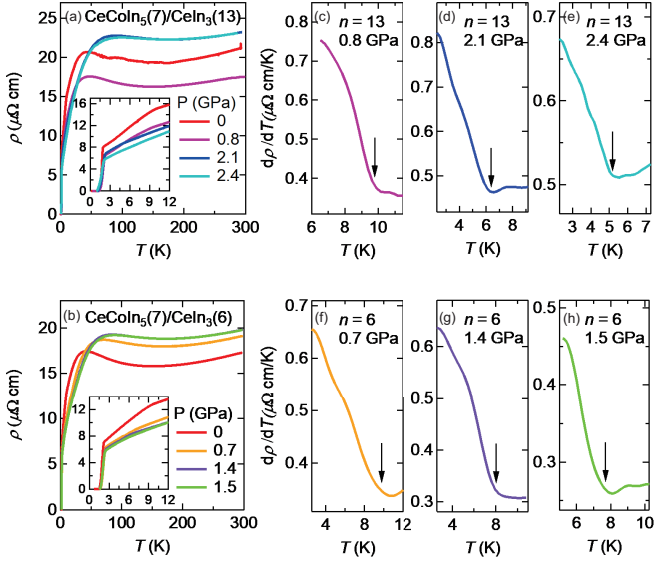}
		\caption{(a), (e) Temperature dependence of the resistivity $\rho(T)$ under pressure for CeCoIn$_5$(7)/CeIn$_3$($n$)  for (a) $n$=13 and (e) $n$=6. Inset: $\rho(T)$ at low temperatures. (b)-(d) and (f)-(h) show the temperature derivative of the resistivity, $d\rho(T)/dT$, as a function of temperature under pressure for $n$=13 and $n$=6, respectively.  The arrows indicate the N\'{e}el temperature $T_N$.}
	\end{center}
\end{figure}

The pressure dependence of the superconducting and magnetic properties provide crucial information on the mutual interaction between superconductivity and magnetism through the interface.  Figures\,5(a) and 5(b) and their insets show the temperature dependence of $\rho(T)$ under pressure for CeCoIn$_5$(7)/CeIn$_3$($n$) for $n$=13 and 6, respectively.   With the application of pressure, the temperature at which $\rho(T)$ shows its maximum increases due to the enhancement of the $c$-$f$ hybridization\cite{Nakajima2007}. 
 As shown in the insets, both superlattices undergo a superconducting transition under pressure.   Figures\,5(c)-5(e) and 5(f)-5(h) show $d\rho(T)/dT$ under pressure for $n$=13 and 6, respectively.  Clear kink structure associated with the AFM transition can be seen in the data. 

\begin{figure}[t]
	\begin{center}
		\includegraphics[width=1\linewidth]{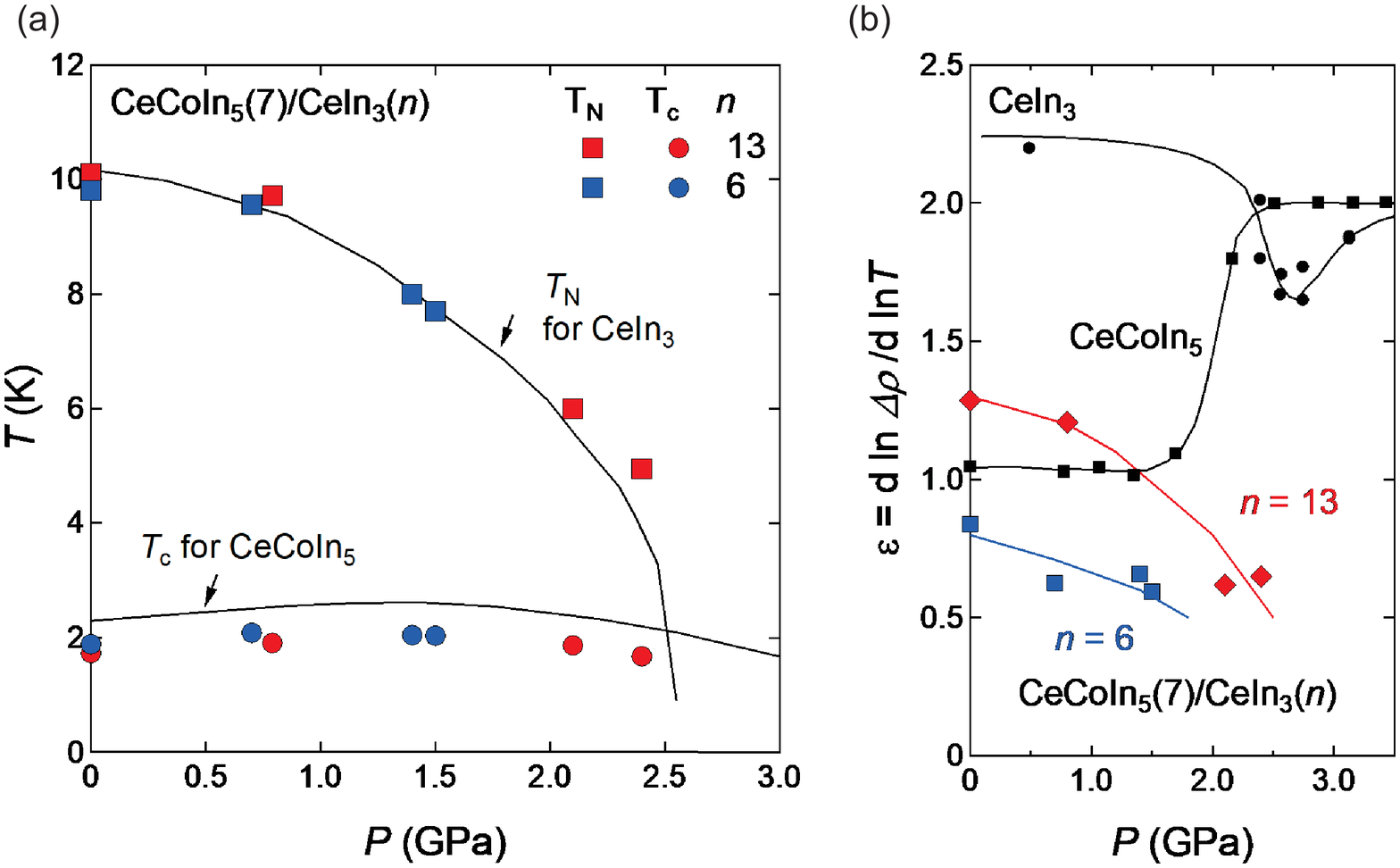}
		\caption{(a) Pressure dependence of $T_N$ and $T_c$ of CeCoIn$_5$(7)/CeIn$_3$($n$) superlattices for $n$=13 and 6.  For comparison,  $T_N$ of CeIn$_3$ and $T_c$ of CeCoIn$_5$ single crystals are shown by solid lines. (b) Pressure dependence of the exponent $\varepsilon$ in $\rho(T)=\rho_0+AT^\varepsilon$, obtained from $d\ln\Delta\rho/d\ln T$ ($\Delta\rho=\rho(T)-\rho_0$), for the CeCoIn$_5$(7)/CeIn$_3$($n$) superlattices for $n$=13 and 6. For comparison, $\varepsilon$ for bulk CeIn$_3$ and CeCoIn$_5$ single crystals is shown. }
	\end{center}
\end{figure}

Figure\,6(a) depicts the pressure dependence of $T_N$ and $T_c$ for CeCoIn$_5$(7)/CeIn$_3$($n$) superlattices for $n$=6 and 13. With applying pressure,  $T_N$ decreases rapidly. For comparison, $T_N$ of a bulk single crystal CeIn$_3$  is also shown by the solid line \cite{Mathur1998}. The pressure dependence of $T_N$ of both superlattices are very similar to that of the bulk CeIn$_3$ single crystal.  In bulk CeIn$_3$ crystal, the AFM QCP is located at $p_c\approx 2.6$\,GPa.  It is natural to expect, therefore,  that the AFM QCP of the superlattices is close to 2.6\,GPa.  Thus, at 2.4\,GPa, the superlattices are in the vicinity of the AFM QCP.    This is supported by the temperature dependence of the resistivity under pressure. The resistivity can be fitted as 
\begin{equation}
\rho(T)=\rho_0+AT^{\varepsilon}.
\end{equation} 
Figure\,6(b) shows the pressure dependence of $\varepsilon$ obtained from $d \ln\Delta \rho/d \ln T$, where $\Delta\rho=\rho(T)-\rho_0$.   The magnitude of $\varepsilon$ decreases with pressure. In bulk CeIn$_3$ single crystal, $\varepsilon$ decreases with pressure and exhibits a minimum at the AFM QCP\cite{Mathur1998,Knebel2001}.  
On the other hand, applying pressure to CeCoIn$_5$ leads to an increase of $\varepsilon$, which is attributed to the 
 suppression of the non-Fermi liquid behavior, $\rho(T)\propto T$, and the development of a Fermi liquid state with its characteristic $\rho(T)\propto T^2$ dependence\cite{Sidorov2002,Nakajima2007}. 
Therefore, the reduction of $\varepsilon$ with pressure arises from the CeIn$_3$ BLs, indicating that the CeIn$_3$ BLs approach the AFM QCP.

\begin{figure}[t]
	\begin{center}
		\includegraphics[width=0.8\linewidth]{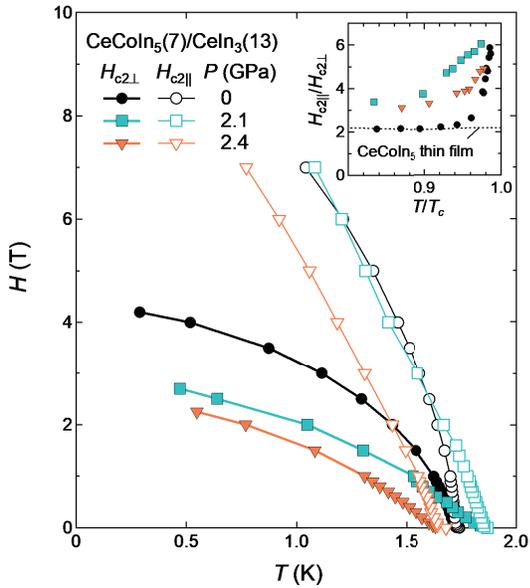}
		\caption{Temperature dependence of upper critical fields in magnetic fields parallel ($H_{c2\parallel}$, open symbols) and perpendicular ($H_{c2\perp}$, closed symbols) to the $ab$-plane for CeCoIn$_5$(7)/CeIn$_3$(13) superlattice at ambient pressure and at 2.1 and 2.4\,GPa. The inset shows anisotropy of the upper critical field, $H_{c2\parallel}/H_{c2\perp}$.  The data of CeCoIn$_5$ thin film at ambient pressure is shown by dotted line. }
	\end{center}
\end{figure}

As shown in Fig.\,6(a), $T_c$ increases, peaks at $\sim$ 1.8\,GPa, and then decreases when applying pressure.  This pressure dependence bears resemblance to that of CeCoIn$_5$ bulk single crystals \cite{Sidorov2002}.    An  analysis of the upper critical field provides important information about the superconductivity of CeCoIn$_5$ BLs.  Figure\,7 depicts the temperature dependence of the upper critical field determined by the midpoint of the resistive transition in a magnetic field $\bm{H}$ applied parallel ($H_{c2\parallel}$) and perpendicular ($H_{c2\perp}$) to the layers.  The inset of Fig.\,7 shows the anisotropy of the upper critical fields $H_{c2\parallel}/H_{c2\perp}$ at ambient pressure.  The anisotropy diverges  on approaching $T_c$.  This is in sharp contrast to the CeCoIn$_5$ thin film, whose anisotropy is nearly temperature independent up to $T_c$.  The observed diverging anisotropy indicates that the superconducting electrons are confined in the 2D CeCoIn$_5$ BLs.  In fact, in  2D superconductivity, $H_{c2\parallel}$ is limited by Pauli paramagnetic pair breaking and increases as $\sqrt{T_c-T}$, while $H_{c2\perp}$ increases as $T_c-T$ due to the orbital pair breaking near $T_c$ \cite{Mizukami2011}.  Moreover, the thickness of the CeCoIn$_5$ BL is comparable to the coherence length perpendicular to the layer, $\xi_c\sim 4$\,nm.  Thus each 7-UCT CeCoIn$_5$ BL effectively behaves as a 2D superconductor. 
 
 \section{Discussion}
 
 \begin{figure}[t]
	\begin{center}
		\includegraphics[width=1\linewidth]{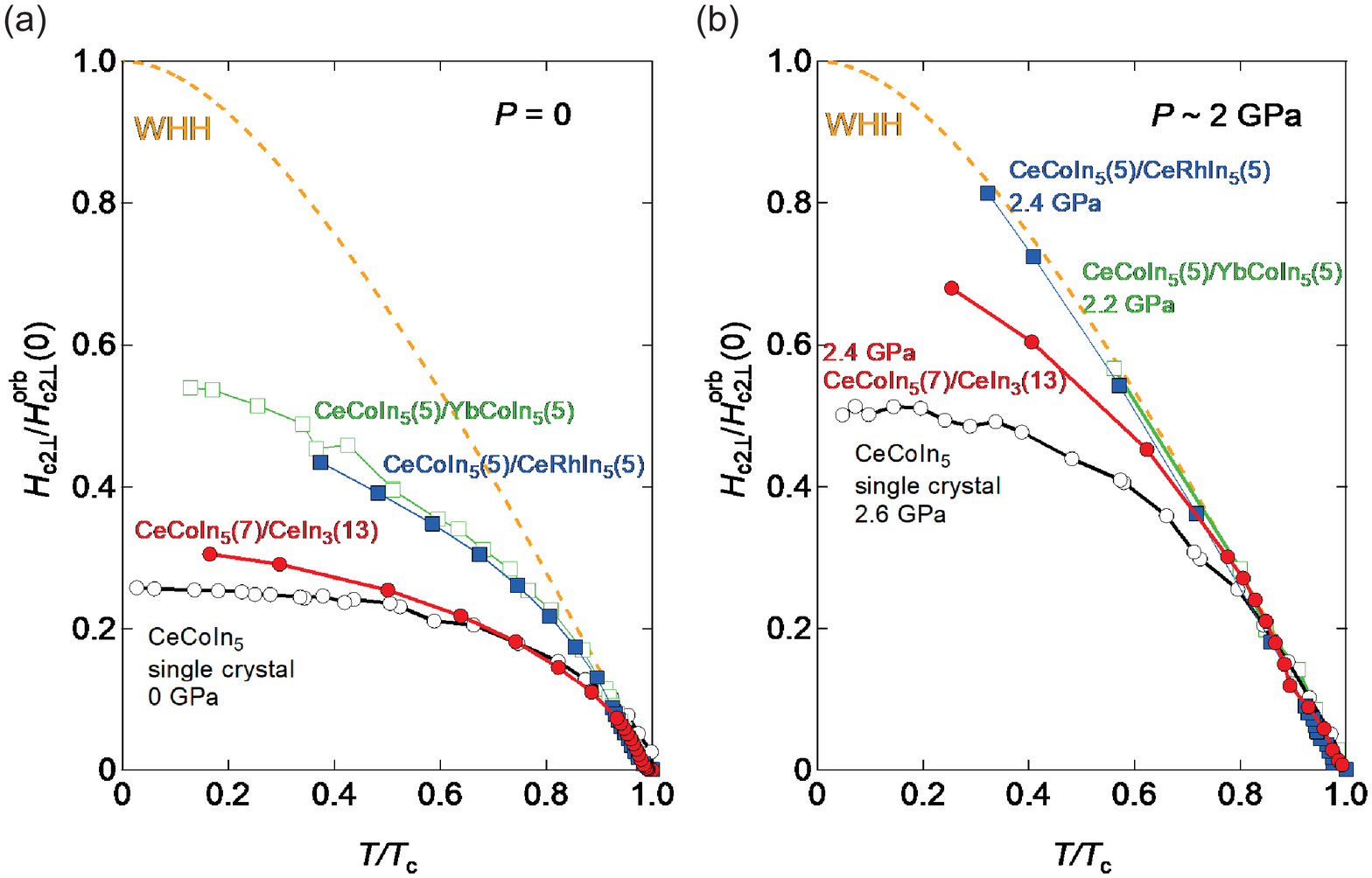}
		\caption{(a) Upper critical field in perpendicular field normalized by the orbital limiting upper critical field, $H_{c2\perp}/H_{c2\perp}^{orb}(0)$,  plotted as a function of $T/T_c$ (a) at ambient pressure and (b) under pressure about 2\,GPa for CeCoIn$_5$(7)/CeIn$_3$(13) superlattices.  For comparison,  $H_{c2\perp}/H_{c2\perp}^{orb}(0)$ for bulk CeCoIn$_5$ single crystal, CeCoIn$_5$(5)/YbCoIn$_5$(5) and CeCoIn$_5$(5)/CeRhIn$_5$(5) are shown. Orange dotted lines represent the WHH curve, which is upper critical field  for purely orbital limiting.}
	\end{center}
\end{figure}

It has been revealed that the temperature dependence of  $H_{c2\perp}$  provides crucial information about the impact of the interface on the superconductivity in  CeCoIn$_5$ BLs.   In particular, the modification of the Pauli paramagnetic effect in the superlattice, which dominates the pair breaking in bulk CeCoIn$_5$ single crystals,  gives valuable clues\cite{Goh2012,Shimozawa2014,Naritsuka2017,Naritsuka2018}. 
 Figure\,8(a) and 8(b) depict the $T$ dependence of the $H_{c2\perp}$ of CeCoIn$_5$(7)/CeIn$_3$(13) superlattice, normalized by the orbital-limited  upper critical field at zero temperature, $H_{c2\perp}^{orb}(0)$, which is obtained from the Werthamer-Helfand-Hohenberg (WHH) formula,  $H_{c2\perp}^{orb}(0)=-0.69T_c(dH_{c2\perp}/dT)_{T_c}$\cite{WHH}.  In Figs.\,8(a) and 8(b),  two extreme cases are also included; the WHH curve with no Pauli pair-breaking and $H_{c2}/H_{c2\perp}^{orb}(0)$ for bulk CeCoIn$_5$ single crystal \cite{Tayama2002}.   For comparison, $H_{c2\perp}^{orb}(0)$ for CeCoIn$_5$/YbCoIn$_5$ and CeCoIn$_5$/CeRhIn$_5$ are also shown\cite{Mizukami2011, Naritsuka2018}. 

At ambient pressure, $H_{c2\perp}/H_{c2\perp}^{orb}(0)$ of CeCoIn$_5$/YbCoIn$_5$ and CeCoIn$_5$/CeRhIn$_5$ are strongly enhanced from that of CeCoIn$_5$ bulk single crystals, indicating the suppression of the Pauli paramagnetic pair-breaking effect. However, it has been pointed out that the mechanisms of this suppression in these two systems are essentially  different.   In CeCoIn$_5$/YbCoIn$_5$,  the enhancement of $H_{c2\perp}/H_{c2\perp}^{orb}(0)$ is caused by the local inversion symmetry breaking at the interface \cite{Goh2012, Maruyama2012}.  The asymmetry of the potential perpendicular to the 2D plane of the superlattice, $\nabla V \parallel$[001], induces the Rashba spin-orbit interaction $\alpha_R= {\bm g}({\bm k})\cdot {\bm \sigma}\propto ({\bm k} \times \nabla V)\cdot {\bm \sigma}$, where ${\bm g}({\bm k})=(k_y,-k_x,0)/k_F$, $k_F$ and ${\bm \sigma}$ are the Fermi wave number and the Pauli matrices, respectively. The Rashba spin-orbit interaction splits the Fermi surface into two sheets with different spin textures\cite{Bauer2012}.  The energy splitting is given by $\alpha_R$, and the spin direction is tilted into the 2D plane, rotating clockwise on one sheet and anticlockwise on the other.  
When the Rashba splitting exceeds the superconducting gap energy ($\alpha_R>\Delta$), the superconducting state is dramatically modified\cite{Maruyama2012,Bauer2012,Fujimoto2007}.  In particular, when the magnetic field is applied perpendicular to the 2D plane, the magnetic field does not couple to the spins, leading to a suppression of the Pauli pair-breaking effect.  
At $p$=2.2\,GPa, $H_{c2\perp}/H_{c2\perp}^{orb}(0)$ of CeCoIn$_5$/YbCoIn$_5$ nearly  coincides with the WHH curve. This indicates that $H_{c2\perp}$ is dominated by the orbital pair breaking most likely due to the suppression of the Pauli paramagnetic pair-breaking effect by the  Rashba splitting.

On the other hand, in CeCoIn$_5$/CeRhIn$_5$ superlattices, it has been shown that the effect of the local inversion symmetry breaking on $H_{c2\perp}$  is  less important  compared with CeCoIn$_5$/YbCoIn$_5$ \cite{Naritsuka2018}.   It has been proposed that magnetic fluctuations (paramagnons) in CeRhIn$_5$ BLs injected through the interface dramatically enhance the force binding superconducting electron pairs in CeCoIn$_5$ BLs, leading to the enhancement of $\Delta$.   As a result, the Pauli limiting field $H_{c2\perp}^{Pauli}(=\sqrt{2}\Delta/g\mu_B$) is enhanced, where  $g$ is the $g$-factor of the electrons.  This increases  the relative importance of the orbital pair-breaking effect, giving rise to the enhancement of $H_{c2\perp}/H_{c2\perp}^{orb}(0)$ \cite{Naritsuka2018}.   At $p$=2.1\,GPa, which is close to the AFM QCP of CeRhIn$_5$ BLs,   $H_{c2\perp}/H_{c2\perp}^{orb}(0)$ nearly  coincides with the WHH curve.   This has been attributed to the enhanced Pauli limiting field  that well exceeds the orbital limiting field ($H_{c2\perp}^{Pauli}\gg H_{c2\perp}^{orb}$) \cite{Naritsuka2018}. 

In contrast to CeCoIn$_5$/YbCoIn$_5$ and  CeCoIn$_5$/CeRhIn$_5$,  $H_{c2\perp}/H_{c2\perp}^{orb}(0)$ is only slightly enhanced in CeCoIn$_5$(7)/CeIn$_3$(13) superlattice at ambient pressure from that of bulk CeCoIn$_5$ single crystal. This indicates that $H_{c2\perp}$ is dominated by Pauli paramagnetic effect, i.e. $H_{c2\perp}\approx H_{c2\perp}^{Pauli}\ll H_{c2\perp}^{orb}$. 
 This implies that the effect of local inversion symmetry breaking  on the superconductivity in CeCoIn$_5$/CeIn$_3$ is weak compared with CeCoIn$_5$/YbCoIn$_5$.   The local inversion symmetry is broken for the CeCoIn$_5$/YbCoIn$_5$ on the CoIn-layer while it is broken on the Ce layer for CeCoIn$_5$/CeIn$_3$ and CeCoIn$_5$/CeRhIn$_5$. Therefore, the present results suggest that the inversion symmetry breaking on the CoIn-layer induces a larger local electric field gradient.   Moreover, superconducting electrons in CeCoIn$_5$ BLs are not strongly influenced by the AFM order in CeIn$_3$ BLs compared with CeCoIn$_5$/CeRhIn$_5$. 

\begin{figure}[t]
	\begin{center}
		\includegraphics[width=0.7\linewidth]{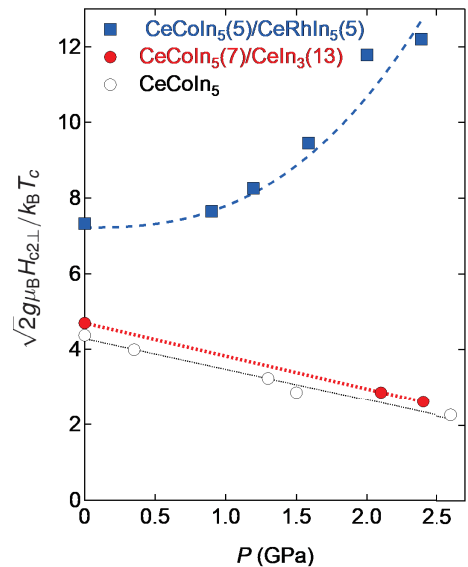}
		\caption{Pressure dependence of $q=\sqrt{2}g\mu_BH_{c2\perp}/k_BT_c\approx 2\Delta/k_BT_c$ for CeCoIn$_5$(7)/CeIn$_3$(13) superlattice.  For comparison, $q$ of bulk CeCoIn$_5$ single crystal and CeCoIn$_5$(5)/CeRhIn$_5$(5) are plotted.}
	\end{center}
\end{figure}

When superconductivity is dominated by the Pauli-limiting effect ($H_{c2\perp}\approx H_{c2\perp}^{Pauli}$),  $2\Delta/k_BT_c$ is estimated as
\begin{equation}
\frac{2\Delta}{k_B T_c}\approx \sqrt{2} \frac{g\mu_BH_{c2\perp}}{k_BT_c}.
\end{equation}
Figure\,9 depicts the pressure dependence of $q=\sqrt{2}g\mu_BH_{c2\perp}/k_BT_c$ for  CeCoIn$_5$/CeRhIn$_5$ and CeCoIn$_5$/CeIn$_3$, along with $q$ for bulk CeCoIn$_5$ single crystal.  Here $g$=2 is assumed.  We note that $q=4.2$ of the bulk CeCoIn$_5$ is smaller than the value determined by the specific heat measurements $2\Delta/k_BT_c\approx$ 6 \cite{Petrovic2001}, but is larger than the BCS value of $q=3.54$, which is  consistent with the strong coupling  superconductivity.   The increase of $q$ with pressure in CeCoIn$_5$/CeRhIn$_5$ implies the increase of  $2\Delta/k_BT_c$.  This  increase has been attributed to an enhancement of the force binding superconducting electron pairs. In spin fluctuation mediated superconductors, the pairing interaction is mainly provided by high-energy fluctuations while low-energy fluctuations act as pair breaking.  In this case, an increase of $2\Delta/k_BT_c$ occurs without accompanying a large enhancement of $T_c$, which is consistent with the results of  CeCoIn$_5$/CeRhIn$_5$ \cite{Naritsuka2018}. Thus, the critical AFM fluctuations that develop in CeRhIn$_5$ BLs near the QCP are injected into the CeCoIn$_5$ BLs through the interface and strongly enhance the pairing interaction in CeCoIn$_5$ BLs. 

In stark contrast to CeCoIn$_5$/CeRhIn$_5$ superlattices, $q$ decreases with pressure in bulk CeCoIn$_5$ single crystal.  This implies that the pairing interaction is weakened with applying pressure, which is consistent with the fact that the pressure moves the system away from the QCP of CeCoIn$_5$.  The reduction of $2\Delta/k_BT_c$ with pressure in bulk CeCoIn$_5$ single crystals is confirmed by the jump of the specific heat at $T_c$ \cite{Knebel2004}.  It should be stressed that the pressure dependence of  $q$ in CeCoIn$_5$(7)/CeIn$_3$(13) is very similar to that of bulk CeCoIn$_5$. This strongly indicates that the pairing interactions in CeCoIn$_5$ BLs are barely influenced by AFM fluctuations injected from the adjacent CeIn$_3$ BLs through the interface even when CeIn$_3$ BLs are located near the AFM QCP.    

The most salient feature in the CeCoIn$_5$/CeIn$_3$ superlattices is that the superconductivity of CeCoIn$_5$ BLs is little affected by the critical AFM fluctuations  in CeIn$_3$ BLs, despite the fact that AFM fluctuations are injected from the adjacent CeIn$_3$ BLs into CeCoIn$_5$ BLs, as evidenced by the AFM order in CeCoIn$_5$/CeIn$_3$  demonstrating  that different CeIn$_3$ BLs are magnetically coupled  by the RKKY interaction through adjacent CeCoIn$_5$ BLs. 
Even in the vicinity to the AFM QCP of the CeIn$_3$ BLs, the superconducting state in the CeCoIn$_5$ BLs is very similar to that of CeCoIn$_5$ bulk single crystals.
  This indicates that the AFM fluctuations injected from CeIn$_3$ BLs do not help to enhance the force binding the superconducting electron pairs in CeCoIn$_5$ BLs. 
    
     This is in stark contrast to CeCoIn$_5$/CeRhIn$_5$, in which the pairing force in CeCoIn$_5$ BL is strongly enhanced by the AFM fluctuations in CeRhIn$_5$ BLs\cite{Naritsuka2018},  although the CeRhIn$_5$ BLs are magnetically only weakly coupled through CeCoIn$_5$ BLs.    We point out that these contrasting behaviors can be attributed to the differences of the magnetic and electronic properties of CeRhIn$_5$ and CeIn$_3$.   The magnetic wave vector in the ordered phase of CeIn$_3$ is commensurate $\bm{q}_0$=(0.5,0.5,0.5)\cite{Benoit1980}.  The evolution of the ordered moment below $T_N$ is consistent with mean field theory. On the other hand, the magnetic wave vector in the ordered phase of CeRhIn$_5$ is  incommensurate $\bm{q}_0$=(0.5,0.5,0.297)\cite{Bao2000}.  The evolution of the ordered moment below $T_N$ deviates from mean field behavior,  likely due to 2D fluctuations.      In CeCoIn$_5$,  AFM fluctuations with wave vector  $\bm{q}_f$=(0.45, 0.45, 0.5) are dominant\cite{Raymond2015}.   Thus, the $c$ axis component of $\bm{q}_f$ in  CeCoIn$_5$ is commensurate and  has the same value as that of $\bm{q}_0$ in CeIn$_3$.  On the other hand, the $c$ axis component of $\bm{q}_0$ in CeRhIn$_5$ is incommensurate and very different from that of $\bm{q}_0$ in CeCoIn$_5$. 
     
     The equality between the $c$ axis component of $\bm{q}_f$ in  CeCoIn$_5$ and $\bm{q}_0$ in CeIn$_3$ would explain why the magnetic coupling between CeIn$_3$ BLs through CeCoIn$_5$ BL is stronger than that between CeRhIn$_5$ BLs.  Thus, AFM order is formed in CeCoIn$_5$(7)/CeIn$_3$($n$) even for small $n$, for which the  
    AFM order has already vanished in   CeCoIn$_5$($n$)/CeRhIn$_5$($n$).
In magnetically mediated superconductors, the pairing interaction is expected to be strongly wave number dependent.  Considering that the quasi-2D Fermi surface of CeCoIn$_5$ bears a close resemblance to that of CeRhIn$_5$ and the superconducting pairing state of both compounds is $d_{x^2-y^2}$\cite{Park2008},  it is likely that the pairing interaction in both compounds has 2D character and peaks around the same wave number.  Furthermore, it has been assumed that 2D magnetic fluctuations are strong in CeRhIn$_5$. Thus,  superconductivity in the CeCoIn$_5$ BLs of CeCoIn$_5$($n$)/CeRhIn$_5$($n$) is strongly influenced.  
     On the other hand, AFM fluctuations having 3D character in CeIn$_3$  may not play an important role for the pairing interaction in  CeCoIn$_5$, resulting in little change of the superconductivity in CeCoIn$_5$/CeIn$_3$.\\

\section{Summary}

A state-of-the-art MBE technique has enabled us to fabricate superlattices consisting of different heavy-fermion compounds. These Kondo superlattices provide a unique opportunity to study the mutual interaction between unconventional superconductivity and magnetic order through the atomic interface.   In hybrid Kondo superlattice CeCoIn$_5$/CeIn$_3$, the superconductivity in CeCoIn$_5$ BLs and AFM order in CeIn$_3$ BLs coexist in spatially separated layers. We find that each CeIn$_3$ BL is magnetically coupled by the RKKY interaction through adjacent CeCoIn$_5$ BLs.  An  analysis of the upper critical field under pressure reveals that the  superconductivity in CeCoIn$_5$ BLs is little influenced  even in the presence of abundant AFM fluctuations in the vicinity of the AFM QCP of adjacent CeIn$_3$ BLs.   Thus, although the AFM fluctuations are injected  to the CeCoIn$_5$ BLs from the CeIn$_3$ BLs through the interfaces,  they barely influence the force binding superconducting electron pairs.  This is in sharp contrast to CeCoIn$_5$/CeRhIn$_5$, in which the superconductivity in the CeCoIn$_5$ BLs are strongly influenced by quantum critical AFM fluctuations in CeRhIn$_5$ BLs.  
 
   It has been widely believed that 2D AFM fluctuations are important for the pairing interaction in CeCoIn$_5$.  However, direct evidence was lacking. The present results provide strong support that 2D AFM fluctuations are essentially important for the unconventional superconductivity in CeCoIn$_5$.

\section*{Acknowledgements}
We thank K. Ishida, H. Kontani, and Y. Yanase for fruitful discussions. This work was supported by Grants-in-Aid for Scientific Research (KAKENHI) (Nos. 25220710, 15H02014, 15H02106, 17K18753, 18H05227, 18J10553), and on Innovative Areas `Topological Material Science' (No. JP15H05852) and `3D Active-Site Science' (No. 26105004) from Japan Society for the Promotion of Science (JSPS). M. N.  also acknowledges support from a JSPS Fellows.

\end{document}